\documentclass[journal]{IEEEtran}

\usepackage{amsmath,amsfonts,bm,amssymb,psfrag,ifthen,color,subfigure}
\usepackage{mathrsfs}
\usepackage[justification=centering,font=footnotesize]{caption}
\usepackage{todonotes}
\usepackage[bottom]{footmisc}

\allowdisplaybreaks[4]

\usepackage[dvips]{epsfig}
\usepackage[dvips]{graphicx}
\usepackage{amsmath,amsfonts,bm,amssymb,psfrag,ifthen,color,subfigure}
\usepackage{algpseudocode}
\usepackage{algorithm}
\usepackage{algorithmicx}
\usepackage{ifthen}
\usepackage{setspace}
\usepackage{cite}
\usepackage{url}
\usepackage{longtable}
\usepackage{stfloats}
\usepackage{graphics,booktabs,color,subfigure}
\usepackage{float}
\usepackage{diagbox}
\usepackage{threeparttable}
\usepackage{tabularx}
\usepackage{hyperref} 
\usepackage{caption}
\usepackage{listings}

\floatname{algorithm}{Algorithm}

\begin{document}
\title{Key Issues in Wireless Transmission for NTN-Assisted Internet of Things}
\author{Chenhao~Qi, Jing Wang, Leyi Lyu, Lei Tan, Jinming Zhang, and Geoffrey Ye Li
\thanks{Chenhao~Qi, Jing Wang, Leyi Lyu, Lei Tan and Jinming Zhang are with the School of Information Science and Engineering, Southeast University, China. }
\thanks{Geoffrey Ye Li is with the Department of Electrical and Electronic Engineering, Imperial College London, U.K.}
\thanks{This work was supported in part by National Key Research and Development Program of China under Grant 2021YFB2900404.} 
}

\markboth{Accepted by IEEE Internet of Things Magazine}
{}

\maketitle
\IEEEpeerreviewmaketitle
	
\begin{abstract}
Non-terrestrial networks (NTNs) have become appealing resolutions for seamless coverage in the next-generation wireless transmission, where a large number of Internet of Things (IoT) devices diversely distributed can be efficiently served.	The explosively growing number of IoT devices brings a new challenge for massive connection. The long-distance wireless signal propagation in NTNs leads to severe path loss and large latency, where the accurate acquisition of channel state information (CSI) is another challenge, especially for fast-moving non-terrestrial base stations (NTBSs). Moreover, the scarcity of on-board resources of NTBSs is also a challenge for resource allocation. To this end, we investigate three key issues, where the existing schemes and emerging resolutions for these three key issues have been comprehensively presented. The first issue is to enable the massive connection by designing random access to establish the wireless link and multiple access to transmit data streams. The second issue is to accurately acquire CSI in various channel conditions by channel estimation and beam training, where orthogonal time frequency space modulation and dynamic codebooks are on focus. The third issue is to efficiently allocate the wireless resources, including power allocation, spectrum sharing, beam hopping, and beamforming. At the end of this article, some future research topics are identified.		
\end{abstract}


\section{Introduction}
\IEEEPARstart{W}{ith} the fast development of wireless communications, Internet of Things (IoT) have revolutionized to enable massive connection of diverse devices, including vehicles, sensors, wearable terminals, and etc. However, for the IoT communications-based terrestrial networks (TNs), the wireless service for massive connection may not be available in remote regions, such as deserts, oceans and mountains. Besides, the increasing number and growing demand of IoT devices call for  wireless service that cannot {be} sufficiently provided by current TNs~\cite{siot}.

	
As a complement for TNs, the non-terrestrial networks (NTNs) including satellite networks and air-based networks have attracted extensive interest from both industry and academia. The assist of NTNs for traditional wireless communications can provide full-time and three-dimensional broadband access for ubiquitous coverage on the earth. In addition, the non-terrestrial base stations (NTBSs) can provide more resources for TNs by offering supplementary traffic for the IoT devices. However, the explosively growing number of IoT devices brings a new challenge for massive connection. The long-distance wireless signal propagation in NTNs leads to severe path loss and large latency, where the accurate acquisition of channel state information (CSI) is another challenge, especially for fast-moving NTBSs. In addition, the scarcity of on-board resources of NTBSs is also a challenge for resource allocation. Therefore, three issues including massive connection, CSI acquisition and resource allocation will be addressed for the NTN-assisted IoT communications in this article. {Note that different from the existing works, this article focuses on the issues for the physical-layer wireless transmission.}   

To enable massive connection for a large number of IoT devices and overcome the shortage of orthogonal preambles, random access (RA) is adopted to establish the wireless link in the first stage, while non-orthogonal multiple access (NOMA) instead of orthogonal multiple access (OMA) is adopted for data transmission in the second stage. To perform sporadic communications between the NTBSs and massive IoT devices, the RA schemes including grant-based RA, grant-free RA and unsourced RA are efficient {resolutions}. To efficiently utilize the limited {resources} of NTNs, NOMA can be adopted.


	

To acquire the CSI for NTN-assisted IoT communications, channel estimation and beam training are adopted, where large propagation latency and high mobility should be considered. Orthogonal time frequency space (OTFS) modulation can be introduced to deal with the fast mobility of the NTBSs, where the channels can be estimated in the delay-Doppler domain with improved accuracy and the channel sparsity can be exploited. For beam training, the dynamic codebook instead of the static codebook is preferred.  Hierarchical codebooks can be employed to further reduce the training latency. If some prior knowledge such as speed, trajectory, and location is available, beams can be generated with better precision for the training.

To efficiently utilize the limited resource, judicious resource allocation is important. Power allocation can flexibly assign power to different data streams to reduce the total power consumption as well as to improve the performance of NOMA. Spectrum sharing from the frequency domain can alleviate the spectrum scarcity through cognitive radio (CR). Beam hopping (BH) from the time domain can effectively mitigate the co-channel interference among different beams illuminated in the same time slot. Beamforming from the spatial domain can improve both the spectral efficiency and energy efficiency by designing analog beamformer and digital beamformer.



The rest of this article is organized as follows. The architecture of NTN-assisted IoT communications is given in Section~\ref{Architecture}. Three key issues including massive connection, CSI acquisition and resource allocation are addressed in Sections~\ref{Massive},~\ref{CSIacquisition} and \ref{Resource}, respectively. Finally, the conclusion and open issues are provided in Section~\ref{Conclusion}.
	
\section{Architecture of NTN-assisted IoT Communications}\label{Architecture}
\begin{figure}[!t]
		\centering
		\includegraphics[width=1\linewidth]{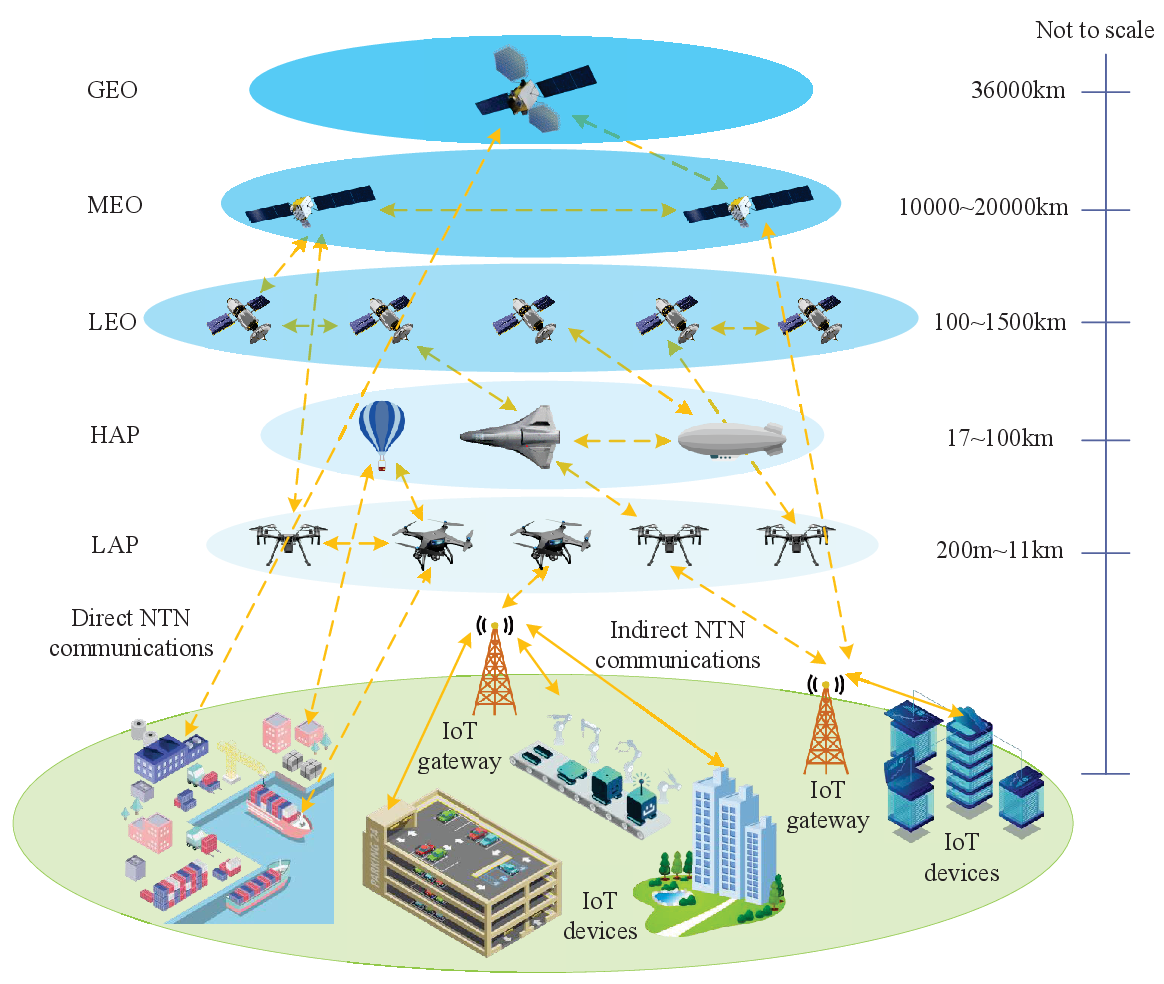}
		\caption{Illustration of NTN-Assisted IoT communications.}
		\label{NTN}
	\end{figure}

The standardization of the NTNs is included in Release 17 of the 3rd generation partnership project (3GPP), where the 5G wireless networks can completely support the NTNs. The NTNs can be generally categorized into satellite networks and air-based networks, depending on whether their operation altitude is higher than 100~km or not. As illustrated in Fig.~\ref{NTN}, the NTBSs in the satellite networks can be mainly classified into geosynchronous Earth orbit (GEO), medium Earth orbit (MEO), and low Earth orbit (LEO) satellites, according to their orbital altitudes. The air-based networks are comprised of high-altitude platform (HAP) networks and low-altitude platform (LAP) networks, where the NTBSs of the former include aerostatic balloon, aerodynamic aerocraft and aerostatic airship, and the NTBSs of the latter include unmanned aerial vehicles~\cite{9210567}.

	The NTN-assisted IoT communications can be mainly divided into two types as follows. 
	
	\begin{itemize}
		\item {\textit{Indirect NTN communications}}: Each IoT device communicates with the NTBSs through at least an IoT gateway on the ground, where the IoT gateways directly communicate with the NTBSs. For indoor applications such as intelligent building, automatic parking system and intelligent assembly, the signal of the NTBSs cannot cover the indoor IoT devices and the signal relay of the IoT gateways is needed. For outdoor IoT devices that are densely distributed, the IoT gateways are also needed to improve the efficiency of the signal transmission of the IoT devices.
		
		
		\item {\textit{Direct NTN communications}}: Each IoT device directly communicates with the NTBSs, where the IoT gateways can be integrated in the NTBSs. For the IoT devices that are sparsely distributed or fast moving, deploying IoT gateways fixed in the ground leads to uneconomical infrastructure and low efficiency in serving the IoT devices. The applications include intelligent logistics and fast delivery of products for remote areas.
		
	\end{itemize}

\begin{figure}[!t]
	\centering
	\includegraphics[width=\linewidth]{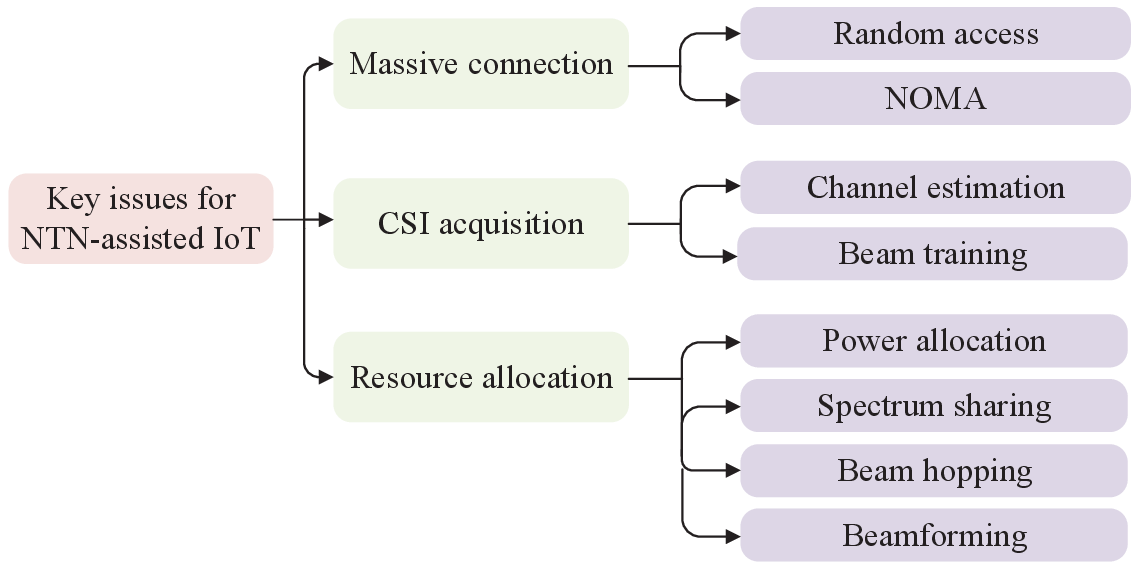}
	\caption{Key issues of NTN-assisted IoT communications.}
	\label{overview}
\end{figure}

The challenges of wireless transmission for NTN-assisted IoT communications mainly come from the large number of IoT devices, various channel conditions and limited wireless resources. To this end, we investigate three key issues, as shown in Fig.~\ref{overview}, where the existing schemes and emerging resolutions for these three key issues will be presented. The first issue is to enable the massive connection by designing RA to establish the wireless link and multiple access (MA) to transmit data streams. The second issue is to accurately acquire CSI in various channel conditions, based on channel estimation and beam training. The third issue is to efficiently allocate wireless resources, including power allocation, spectrum sharing, beam hopping, and beamforming.

\section{Massive Connection}\label{Massive}
\begin{figure*}[!t]
	\centering
	\includegraphics[width=0.8\linewidth]{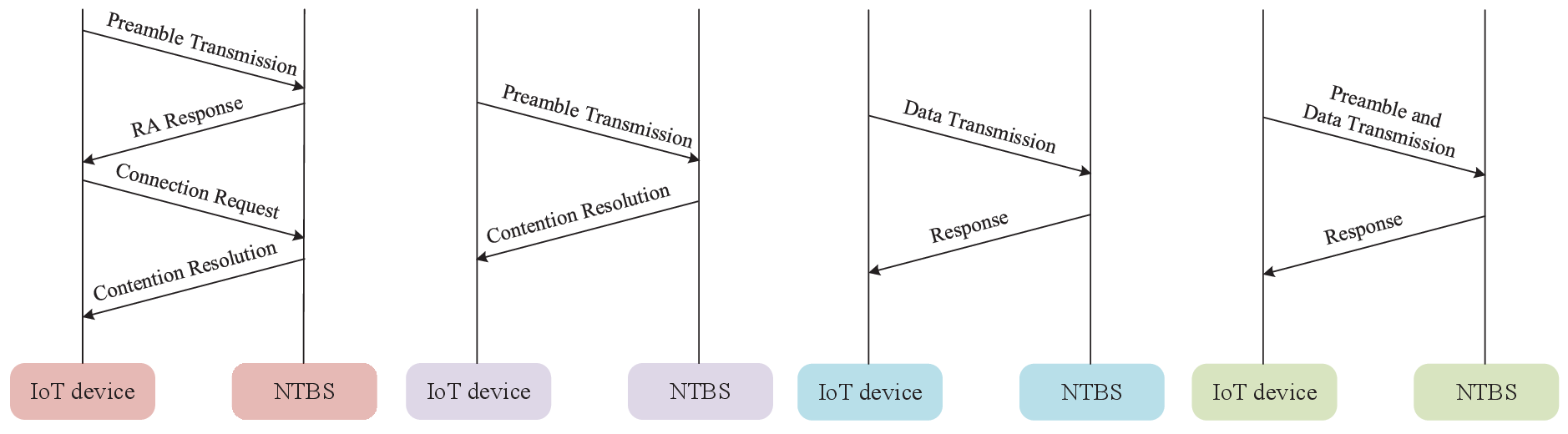}
	\caption{The procedures of grant-based RA, grant-free RA, URA and grant-free NOMA.}
	\label{MA1}
\end{figure*}

The first issue for wireless transmission of NTN-assisted IoT communications is to enable the massive connection for the explosively growing number of IoT devices. Different from terrestrial IoT, the NTN-assisted IoT faces severe transmission latency due to the long distance between the NTBSs and IoT devices, where grant-free techniques are preferred to the grant-based ones. To support the massive IoT devices, RA is adopted to establish the wireless link in the first stage, while NOMA instead of OMA is adopted for data transmission in the second stage. 

\subsection{Random Access}
The RA allows the IoT devices to randomly access the NTNs before the data transmission, without any procedure to grant resources in advance. Therefore, the RA fits for the widely distributed numerous IoT devices, where they can transmit data at different times and from different locations. To save the transmission power, the IoT devices are normally set inactive and are set to active only when there are data to be transmitted to the NTBSs. For the sourced RA (SRA), the NTBSs will obtain the identity (ID) of the active IoT devices followed by their data. For the unsoured RA (URA) that is applicable in scenarios such as distributed sensing, the NTBSs are only interested in the data of the active IoT devices and will not obtain their IDs.  


\begin{itemize}
	\item {\textbf{SRA:}} There are two types of SRA named grant-based RA and grant-free RA. As shown in Fig.~\ref{MA1}, the grant-based RA needs four steps to establish the wireless link, including preamble transmission, RA response, connection request and contention resolution. Considering the transmission latency in the NTNs, we need to reduce the signaling overhead involved in the four steps of the grant-based RA. Then the grant-free RA can be adopted, where active IoT devices directly transmit preambles to the NTBSs without permission and therefore the number of steps can be reduced from four to two. At a certain time instant, the number of active IoT devices is much smaller than that of all the IoT devices, leading the status vector formed by all the IoT devices to be sparse, where the sparse signal processing methods can be employed for active device detection (ADD) in such sporadic communications~\cite{zhangUserActivityDetection2020}. 
	
	\item {\textbf{URA:}} To save the space of the IDs and further improve the transmission efficiency, the active IoT devices in URA do not need to transmit preambles and directly transmit data to the NTBSs, which means the establishment of the wireless link and the data transmission are jointly performed. In fact, all the IoT devices share a common codebook and the NTBS only recovers data blocks instead of the IDs of the active IoT devices. The URA can be further categorized as the stitch-based URA and signature-based URA according to the structure of transmission. For the stitch-based URA, the data block to be transmitted from the active IoT devices can be divided into several sub-blocks. Then the NTBSs detect the sub-blocks jointly to recover the original preambles. For the signature-based URA, the data block from the active IoT devices is comprised of signature sequences and information sequences. The signature sequences are designed for the NTBSs to operate the information sequences and then to perform multiuser detection~\cite{URA}.
\end{itemize}

\subsection{NOMA}\label{powernoma}
Due to the large number of the IoT devices, it is difficult to ensure each IoT device with an orthogonal preamble. Different from OMA where a resource unit can occupied only by an IoT device at a certain time instance, NOMA allows multiple IoT devices to simultaneously use the same resource unit, which substantially improves the efficiency. Especially for the NTNs where the resources are more limited than those in the TNs, NOMA is a good option. {Owing to various requirements of massive IoT devices, 3GPP is planning to enhance the IoT standard to support low cost IoT devices with high connection density and low resource consumption. To this end,  NOMA is considered as a promising solution for massive connectivity in IoT.} Currently, NOMA can be categorized as power domain NOMA and code domain NOMA, as introduced in the following. 
 
\begin{itemize}
	\item {\textbf{Power domain NOMA:}} Exploiting the difference of channel gains among the IoT devices, power domain NOMA  allocates different powers to data streams to be transmitted, where the receiver retrieves the original data streams by successive interference cancellation (SIC). Considering that the difference of channel gains can be weakened due to the long-distance propagation between the NTBSs and IoT devices, some preprocessing methods are needed to make NOMA work {effectively}, e.g., selecting IoT devices that have substantially different channel gains for NOMA. \\
	
	\item {\textbf{Code domain NOMA:}} The code domain NOMA multiplexes the data steams to different IoT devices with different spreading sequences that are typically characterized by sparse, low-density and low inter-correlation properties. Therefore, multiple transmitted signals are overlapped in the same resource unit. Regarding the long distance and low channel gain for NTN-assisted IoT communications, bit-level or symbol-level signal processing such as interleaving and scrambling can be exploited to further improve the transmission performance. 
	
\end{itemize}

By combining the link establishment and data transmission stages together, grant-free NOMA that works {similarly} to the grant-free RA, allows the IoT devices to send the data together with the preamble to the NTBSs, without waiting for the response from the NTBSs~\cite{gao2023grantfree}. Therefore, grant-free NOMA only has one stage with two steps, which can save the power of IoT devices and reduce the round-trip latency of multi-stage signaling. Since the power domain NOMA needs at least two stages, with the first stage to estimate the channel gain and the second stage to allocate different powers based on the estimated channel gain, the power domain NOMA cannot work in the grant-free mode. For the grant-free NOMA, the ADD, CSI acquisition and data extraction can be jointly performed to improve the efficiency of wireless transmission.

\section{CSI Acquisition}\label{CSIacquisition}

\begin{figure}[!t]
	\centering
	\includegraphics[width=0.8\linewidth]{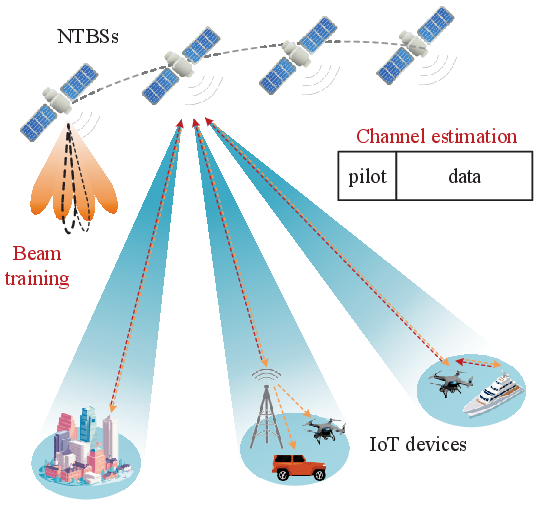}
	\caption{CSI acquisition based on channel estimation or beam training.}
	\label{CSI}
\end{figure}

\begin{figure*}[!t]
	\centering
	\includegraphics[width=0.95\linewidth]{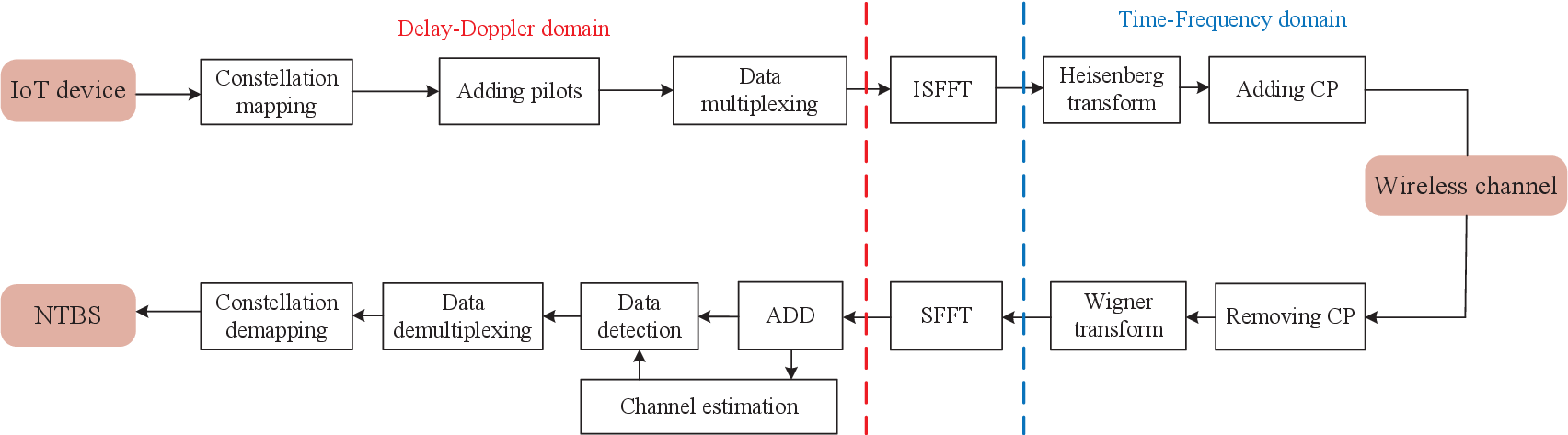}
	\caption{Block diagram of the OTFS modulation.}
	\label{OTFS}
\end{figure*}

The second issue is how to accurately acquire the CSI in various channel conditions, based on channel estimation and beam training. Different from the TN channels, the NTN channels have the following two characteristics:  \textit{i)} The high propagation latency between the NTBSs and IoT devices due to the long transmission distance may lead the CSI outdated when it is acquired. \textit{ii)} The high mobility of some NTBSs such as LEO satellites results in fast time-varying channels and severe Doppler frequency shift. Therefore, in the following, we will investigate the CSI acquisition for NTN-assisted IoT communications from the perspective of channel estimation and beam training, as illustrated in Fig.~\ref{CSI}.

\subsection{Channel Estimation}
Due to the high propagation latency between the NTBSs and IoT devices caused by the long transmission distance, the NTN CSI acquired by the existing terrestrial channel estimation schemes may be outdated. To continuously obtain the instantaneous CSI, a more effective scheme is to design a linear predictor based on the relationship between outdated CSI and instantaneous CSI, where the relationship can be established through analyzing the spatial and temporal correlation of the NTN channel models. Then the factors of the linear predictor can be computed under different criteria such as LS and MMSE. Compared with the existing terrestrial channel estimation schemes, such schemes can improve the accuracy of CSI without additional resource overhead.

{To acquire the CSI for NTN channels with fast moving NTBSs, OTFS modulation as shown in Fig.~\ref{OTFS}, can be adopted. Note that the channels can be estimated in the delay-Doppler domain to improve the accuracy.} Considering massive IoT devices, the pilot-assisted channel estimation requires a large number of pilot {sequences}, which can cause excessive pilot overhead. To reduce the pilot overhead, it {is} helpful to exploit the sparsity in the delay-Doppler domain for channel estimation. To further improve the performance of sparse channel estimation, we can optimize the pilot sequence based on the mutual incoherence property or restrict isometry property under the compressed sensing framework. As another approach to reduce the pilot overhead caused by a large number of IoT devices, grant-free NOMA that does not require mutually orthogonal pilot sequences for IoT devices can be employed, where ADD and channel estimation can be jointly performed with low complexity. By combining the advantage of NOMA to reduce the pilot overhead and that of OTFS against the Doppler effect, OTFS-NOMA can be naturally adopted, where the channels can be simultaneously estimated in the delay-Doppler domain with low pilot overhead based on sparse signal processing methods~\cite{zhouActiveTerminalIdentification2023}.

\subsection{Beam Training}
Different from the TNs, the NTNs have a large coverage for wireless signal, which typically relies on highly directional beamforming to compensate the path loss, especially for the NTNs working at high frequency, e.g., Ku, Ka, Q and V bands. Therefore, we can adopt the beam training, to determine the best beam for IoT devices, and then perform channel estimation using the determined beam in a low dimension. The most straightforward beam training method is beam sweeping, which exhaustively tests all the available beams and then selects the best one. Different from designing the beamformer based on the acquired CSI, beam training is used to acquire the CSI.  

To fast generate beams for beam training, a codebook is generally used. However, different from the TN beam training, the NTN beam training faces the challenges from the large latency and the high-speed movement of NTBSs. Therefore, the dynamic codebook instead of the static codebook is preferred. Note that the codewords in the codebook may be changed in batch according to the time-varying parameters of NTBSs, which only causes a small computational burden for the NTBSs. For example, all the codewords can be multiplied by phase factors that are closely related to the moving speed of the NTBSs. To fast perform beam training so that the latency can be further reduced, hierarchical codebooks including multiple layers of codebooks can be used, where the later-trained layers of codebooks can be dynamically changed according to the beam training results based on the earlier-trained layers of codebooks~\cite{Codebook}.

To generate beams with better precision for beam training, real-time beamforming based on the prior knowledge from the NTBSs and the IoT devices can be considered. For example, the beams might be widened if the speed of NTBSs is higher so that the beams are wide enough to cover the served area in a certain time. If the moving trajectory of NTBSs is available, the beam rotation and attitude adjustment can be made for real-time beamforming. With more information on IoT devices such as their locations or distributions, the beams can be generated with better precision. Since the real-time beam computation might be a burden for NTBSs, we may resort to the machine learning (ML) and use a trained deep neural network (DNN) for real-time beamforming, where the DNN can be trained offline given massive scenarios and parameters. 

After the beam training is finished, the {following} channel estimation can be performed in a low dimension using the determined beams. A small number of pilots or training symbols can be sent and the traditional channel estimation methods including LS or MMSE can be used~\cite{Beamtraining}. 

\section{Resource Allocation}\label{Resource}
\begin{figure}[!t]
	\centering
	\includegraphics[width=1\linewidth]{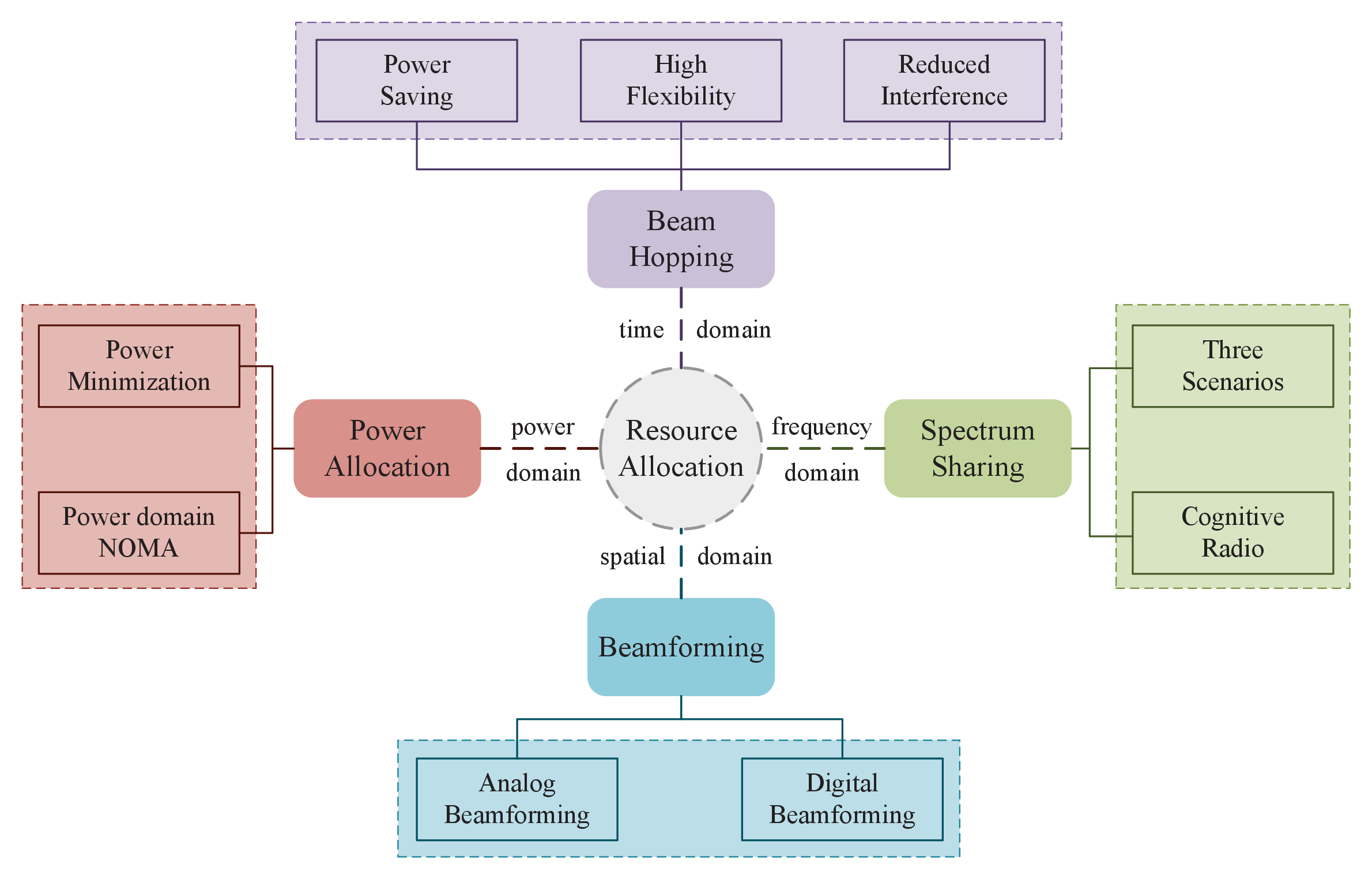}
	\caption{Resource Allocation for NTN-assisted IoT communications.}
	\label{fig5}
\end{figure}

Besides massive connection and CSI acquisition, resource allocation is another key issue, where we consider the resource allocation by investigating power allocation, spectrum sharing, beam hopping and beamforming, from the perspective of power domain, frequency domain, time domain and spatial domain, respectively, as illustrated in Fig.~\ref{fig5}.


\subsection{Power Allocation}\label{powerenergy}

For NTN-assisted IoT communications, power consumption is an important aspect. Due to the difficulty in {frequently} charging the IoT devices or NTBSs, the battery life of the IoT devices or NTBSs is expected to be extended as long as possible. By reducing the transmission power, a long battery life can be achieved. But on the other hand, due to the long transmission distance between the NTBSs and IoT devices with different QoS requirements, certain transmission power should be guaranteed. In this context, we can optimize the power allocation to minimize the transmission power subject to the QoS constraints~\cite{powerconsumption}.

To support downlink transmission from the NTBSs to massive IoT devices, NOMA instead of OMA should be adopted. For the power-domain NOMA,  the lower transmission power of the NTBSs is allocated to data streams for the IoT devices with larger channel gains, and the higher transmission power of the NTBSs is allocated to data streams for the IoT devices with smaller channel gains, so that the SIC can be effectively performed by the IoT devices for desired data recovery. Note that the downlink channel gain can be obtained by channel estimation during the grant-free NOMA uplink transmission from the IoT devices to the NTBSs. Due to the fact that the transmission distance between the NTBSs and some IoT devices might be similarly long, the channel gains of some IoT devices may be similar, which brings the difficulty for the SIC. In this context, it is better to pair the IoT devices into different groups, so that we can use OMA for the wireless transmission among different groups and use NOMA among different IoT devices in the same group. We can pair the IoT devices according to their channel gains, where the channel gain of the IoT devices in the same group should be highly different to facilitate the SIC. Furthermore, we may also take the channel correlation of different IoT devices into {consideration} when pairing them~\cite{powernoma}. For the IoT devices with small channel correlation, we can use the same beam to serve them, which is spatially efficient. Therefore, we may first set a basic requirement for one of the channel gain difference and channel correlation, and then maximize the other one, to jointly consider the NOMA and OMA.

\subsection{Spectrum Sharing}
Owing to the limited available frequency bands, spectral resource for both NTNs and TNs is congested. The large number of IoT devices further exacerbates the severity of the spectrum congestion problem. To alleviate the issue, spectrum sharing is on focus. For the NTN-assisted IoT communications, three scenarios can be considered, including i) spectrum sharing between the NTN systems and the TN systems, ii) spectrum sharing among different NTN systems, iii) spectrum sharing among different TN systems. 

To achieve the spectrum sharing, the CR is a promising technique, since it can detect the available licensed frequency bands of the surrounding environment and give unlicensed system opportunities to use the idle spectrum resources~\cite{cr}. In general, the priorities of the two systems that {share} the same frequency band is not the same. There exists a primary system that has a higher priority to use the licensed spectrum without any constraints, and a secondary system that has a lower priority to utilize the licensed spectrum under certain constraints. According to the constraints imposed on the secondary systems, the way in which the secondary system {utilizes} the spectrum can be divided into two modes: the interweave mode and the underlay mode. In the interweave mode, the secondary system can utilize the idle spectrum only when the primary system does not occupy it. In the underlay mode, the secondary system can utilize the same frequency bands of the primary system with the premise of not significantly affecting the availability of the primary system, where the interference from the secondary system can be constrained by a predefined threshold. {To solve the above problems, interference sensing, power control, using scheduling and beamforming techniques are mainly adopted by the existing literature to suppress the interference from the spectrum sharing~\cite{spectrumsharing}.} 

\subsection{Beam Hopping}
For conventional NTNs without BH, once the beam pattern is given, the beams in the beam pattern will be illuminated continuously without any changes in the time domain. However, only a small number of IoT devices are active at a certain time, which means only a few IoT devices require a short period of data transmission and the continuous beam illumination is unnecessary~\cite{beamhopping}. Since the whole transmission period is normally divided into several time slots, the NTBSs with BH are capable of providing wireless service only in some specific time slots to the active IoT devices by dynamically illuminating the beams where the active IoT devices are located, while the other beams are kept non-illuminated. Compared with the NTBSs without BH, those with BH are superior in the following aspects:
\begin{itemize}
	\item[$\bullet$]
	Since only a small number of the beams are illuminated at each time slot, fewer radio frequency chains and hardware resources are needed, which leads to less power consumption.
	\item[$\bullet$]
	With the high flexibility of BH in the time and space domain, the NTBSs with BH are capable of allocating time slots for beam illumination according to the traffic demand. Therefore, the mismatch between the traffic demand and the traffic supply can be significantly reduced, resulting in better resource utilization.
	\item[$\bullet$]
	The co-channel interference can be substantially reduced by properly selecting the beams to be illuminated, where the beams with some distance instead of adjacent beams are illuminated in priority. 
	
\end{itemize}


\subsection{Beamforming}
According to the location and distribution of the active IoT devices, we can explore the spatial domain and optimize the beamforming of NTBSs by designing the beam direction, beam width and beam shape, which can improve both the spectral efficiency and energy efficiency~\cite{beamforming1}. 

\begin{itemize}
	\item[$\bullet$]
	Beam direction: Due to the long transmission distance between the NTBSs and IoT devices, the propagation of wireless {signals} is severely attenuated. By making the beam point at the IoT devices as accurately as possible, which is also known as the beam alignment, the largest beam gain corresponding to the peak of the beam can effectively compensate for the signal attenuation.  
	
	\item[$\bullet$]
	Beam width: On one hand, the beam width is designed to be large enough so that as more as IoT devices can be covered by the beam. On the other hand, the beam gain should be large enough to ensure the QoS of the IoT devices, which means that  the beam width should not be too large given the constraint on the same transmission power. Therefore, a tradeoff of the beam width is needed. 
	
	
	\item[$\bullet$]
	Beam shape: To improve the signal coverage of the IoT devices, various beam shapes need to be considered. If the IoT devices {located} in two areas that are distantly separated are covered by the same broadcasting data stream, a two-mainlobe beam with each mainlobe pointing at an area can achieve larger beam gain than a single-mainlobe beam with large coverage. To reduce the interference among different IoT devices or different NTBSs when transceiving different data streams, beam nulling needs to be considered, which may also change the beam shape.
	
	
\end{itemize}

In addition to the analog beamforming, digital beamforming should also be considered. Once the analog beamforming is designed, the digital beamforming can be designed to diagonalize the matrix products of the digital beamformer, analog beamformer and the channel matrix, based on LS or MMSE criteria. We can also formulate the optimization problem in terms of digital beamformer to maximize the sum-rate of the NTN-assisted IoT communications, or minimize the transmission power of the NTBSs or IoT devices~\cite{beamforming2}.

\section{Conclusion and Open Issues}\label{Conclusion}
In this article, we have provided an overview on NTN-assisted IoT communications by highlighting three key issues including massive connection, CSI acquisition and resource allocation. For future research, we also identify some open issues as follows.

The architecture of NTN-assisted IoT communications is interesting for future study. Various types of NTBSs and IoT devices may coexist in the same NTN, which makes the massive connection, CSI acquisition and resource allocation complicated. The well-designed heterogeneous NTN architecture can efficiently serve massive IoT devices with different QoS requirements. The coordination among multiple NTBSs can improve the overall system performance. On the other hand, how to eliminate the interference should be carefully addressed.

To reduce the software complexities of real-time computing for resource allocation, AI can be considered. For example, when solving the optimization problem involved in the power allocation, the well-trained DNNs can be used to fast obtain a near-optimal solution. When designing beams in terms of beam direction, width and shape given the prior knowledge such as speed, trajectory or location, AI can be used to fast generate beams. Besides, AI is also attractive for massive connection and CSI acquisition by establishing intelligent connection and low-complexity CSI acquisition algorithms.

To reduce the hardware complexities of NTBSs or IoT
	devices, reconfigurable intelligent surface (RIS) can be considered. The RIS can function as the low-cost antennas for transceivers. The RIS can also work as a signal reflector to cover the areas that the NTN signal cannot reach or to enhance the signal strength of the areas with massive IoT connection. In this context, CSI acquisition as well as resource allocation needs further investigation, since the RIS introduces additional channel links.

To enhance the security of NTN-assisted IoT communications, resource allocation and optimization regarding the performance metrics, such as security rate, security capacity and security outage probability, can be considered. Moreover, signal encryption and artificial noise can be employed to prevent the eavesdropping, and signal processing approaches such as detection, estimation and elimination of various jamming attacks would be interesting for future studies.

\bibliographystyle{IEEE-unsorted}
\bibliographystyle{IEEEtran}

\newpage

\section*{Biographies}
%

\begin{IEEEbiographynophoto}{Chenhao Qi}
	[SM] (qch@seu.edu.cn) is currently a Professor with the School of Information Science and Engineering, Southeast University, Nanjing, China. He also serves as the Head of Jiangsu Multimedia Communication and Sensing Technology Research Center. His research interests include millimeter wave communications, integrated sensing and communication and satellite communications. He received Best Paper Awards from IEEE GLOBECOM in 2019 and IEEE/CIC ICCC in 2022. He has served as an Associate Editor for IEEE Transactions on Communications, IEEE Communications Letters, IEEE Open Journal of the Communications Society, IEEE Open Journal of Vehicular Technology, and China Communications.
\end{IEEEbiographynophoto}

\begin{IEEEbiographynophoto}{Jing Wang}
	received the B.S. degree in communication engineering from Nanjing Normal University, China, in 2019, and the M.S. degree in information and communication engineering from China University of Mining and Technology, Xuzhou, China, in 2022. She is  currently pursuing the Ph.D. degree in signal processing at Southeast University, Nanjing, China.  Her research interests include millimeter wave communications and satellite communications.
\end{IEEEbiographynophoto}

\begin{IEEEbiographynophoto}{Leyi Lyu}
	received the B.S. degree from Nanjing University of Science and Technology, China, in 2022. She is currently pursuing the M.S. degree in signal processing at Southeast University, Nanjing, China. Her research interests include millimeter wave communications and satellite communications.
\end{IEEEbiographynophoto}

\begin{IEEEbiographynophoto}{Lei Tan}
	received the B.S. degree from Dalian University of Technology, China, in 2022. He is currently pursuing the M.S. degree in signal processing at Southeast University, Nanjing, China. His research interests include 5G networks and array signal processing for satellite communications.
\end{IEEEbiographynophoto}

\begin{IEEEbiographynophoto}{Jinming Zhang}
	received the B.S. degree from Southeast University, Nanjing, China, in 2020. He is currently
	pursuing the Ph.D. degree in signal processing at Southeast University. His current research interests include signal processing algorithms for integrated sensing and communication.
\end{IEEEbiographynophoto}

\begin{IEEEbiographynophoto}{Geoffrey Ye Li}
	[F] is currently a Chair Professor at Imperial College London, UK.  He made fundamental contributions to OFDM for wireless communications and introduced deep learning to communications. Dr. Geoffrey Ye Li was elected to IEEE Fellow and IET Fellow for his contributions to signal processing for wireless communications. He won 2024 IEEE Eric E. Sumner Award and several prestigious awards from IEEE Signal Processing, Vehicular Technology, and Communications Societies, including 2019 IEEE ComSoc Edwin Howard Armstrong Achievement Award.
\end{IEEEbiographynophoto}
		
\end{document}